\documentclass[trans]{IEEEtran}
\usepackage[T1]{fontenc}
\usepackage{hyperref}
\usepackage{lettrine}
\usepackage{epsfig}
\usepackage{float}
\usepackage{footnote}
\usepackage{cite}
\usepackage{float}
\usepackage{soul}
\usepackage{amssymb}
\usepackage{amsthm}
\usepackage{amsmath,mathabx}
\usepackage{color}
\usepackage{epstopdf}
\usepackage{tabularx,tabulary}
\usepackage[font=footnotesize]{caption}
\usepackage{graphicx}
\usepackage{caption}
\usepackage{subcaption}
\usepackage{cuted}
\usepackage{dblfloatfix}    
\usepackage{bbm}
\usepackage{algorithm,algorithmic}
\usepackage{multirow}
\usepackage{wrapfig}

\renewenvironment{IEEEbiography}[1]
  {\IEEEbiographynophoto{#1}}
  {\endIEEEbiographynophoto}
\usepackage[table,xcdraw]{xcolor}

\graphicspath{{../}}

\begin{document}
\title{A Software-Defined Opto-Acoustic Network Architecture for Internet of Underwater Things}
\author{
Abdulkadir~Celik,~\IEEEmembership{Member,~IEEE,} Nasir~Saeed,~\IEEEmembership{Senior Member,~IEEE,}
Basem~Shihada,~\IEEEmembership{Senior Member,~IEEE,} Tareq~Y.~Al-Naffouri,~\IEEEmembership{Senior Member,~IEEE,}  and~Mohamed-Slim~Alouini,~\IEEEmembership{Fellow,~IEEE}
\thanks{Authors are with computer, electrical, and mathematical sciences \& engineering (CEMSE) Division at King Abdullah University of Science and Technology (KAUST), Thuwal, KSA. }
}
\maketitle
\begin{abstract}
In this paper, we envision a hybrid opto-acoustic network design for the internet of underwater things (IoUT). Software-defined underwater networking (SDUN) is presented as an enabler of hybridizing benefits of optic and acoustic systems and adapting IoUT nodes to the challenging and dynamically changing underwater environment. We explain inextricably interwoven relations among functionalities of different layers and analyze their impacts on key network attributes.  Network function virtualization (NFV) concept is then introduced to realize application specific cross-layer protocol suites through an NFV management and orchestration system. We finally discuss how SDUN and NFV can slice available network resources as per the diverging service demands of different underwater applications. Such a revolutionary architectural paradigm shift is not only a cure for chronicle underwater networking problems but also a way of smoothly integrating IoUT and IoT ecosystems.
\end{abstract}
\maketitle

\section{Introduction}
\label{sec:intro}
\lettrine{T}{he} internet of things (IoT) is a technological revolution towards integrating physical and digital worlds by interconnecting smart objects to enhance the quality of life in all aspects.  Communication networks undergo the next major change towards fifth-generation (5G) networks to realize this revolution. Lying at the heart of this evolutionary step, network function virtualization (NFV) and software-defined networking (SDN) are recognized as key enablers of a flexible, scalable, agile, and programmable network platform \cite{Yousaf2017NFV}.
 
However, current efforts mostly focus on developing terrestrial IoT solutions without giving sufficient notice on the internet of underwater things (IoUT) applications. Noting that a continuous body of water covers approximately 71\% of the Earth's surface, oceans provide great benefits to the humankind, including climate regulation, food supply, transportation, natural resources, recreation, and medicine. Moreover, oceanic businesses contribute more than 500 billion US dollars to the world economy. Hence, IoUT can mark a new era for scientific, industrial, and military underwater applications, including environmental monitoring, offshore exploration, disaster prevention, tactical surveillance, and assisted navigation.

Nevertheless, 95\% of the oceans are still unexplored because of the peculiarities of underwater communications, networking, and localization \cite{Celik2018Survey}. In this respect, IoUT significantly differs from its terrestrial counterpart (i.e., IoT) in almost every aspect of the network architecture design. First of all, the aquatic medium poses a variety of daunting challenges based on the underlying communication system. Although radio frequency (RF) waves can reach desirable propagation speeds by tolerating turbulent and turbid nature of water, conductivity restricts their transmission range to ten meters and operational bandwidth to 30-300 Hz. For these reasons, underwater RF modules are costly, power-hungry, and bulky due to large antenna size requirements. 

On the other hand, underwater acoustic communication (UAC) is praised and widely used thanks to their several kilometers long communication ranges. However, underwater acoustic networks (UANs) suffer from low achievable rates (10-50 Kbps) and high-variable delay because of limited bandwidth and low propagation speed (1500 m/s) of acoustic signals, respectively. UAC is also susceptible to multi-path fading, Doppler spread, and ambient noise caused by hydrodynamics and vessel traffic \cite{AKYILDIZ2005257}. In return for limited transmission ranges (50-100 m), optical wireless communication (OWC) can provide desirable data rates in the order of Gbps and low latency thanks to the high propagation speed of light in the aqua ($\approx 2.55 \times 10^8$ m/s). The transmission loss of light is primarily characterized by absorption and scattering effects, which constitute range-beamwidth tradeoff and vary with water types and depths. In particular, blue and green wavelengths have shown to be more resilient in clear and coastal waters, respectively.  As UAC and OWC complement each other in many ways, the full benefit can be obtained by hybridizing UANs and underwater optical wireless networks (UONs).

In this paper, we accordingly envision an underwater opto-acoustic network (UOAN) architecture for IoUT ecosystem. For the sake of clarity, UOAN is first presented in a classical layered fashion to describe underlying network infrastructure and identify network functions associated with each layer. Thereafter, we provide recent advances and present challenges by pointing out inextricably interwoven relations among different network functions. Then, we discuss how NFV and SDN can be used to uniformly and coherently orchestrate various network functions and resources across the multiple layers of the UOAN architecture. We finally conclude the paper with prospects of UOANs for IoUT applications.
 \begin{figure*}[!t]
\begin{center}
\includegraphics[width=0.99 \textwidth]{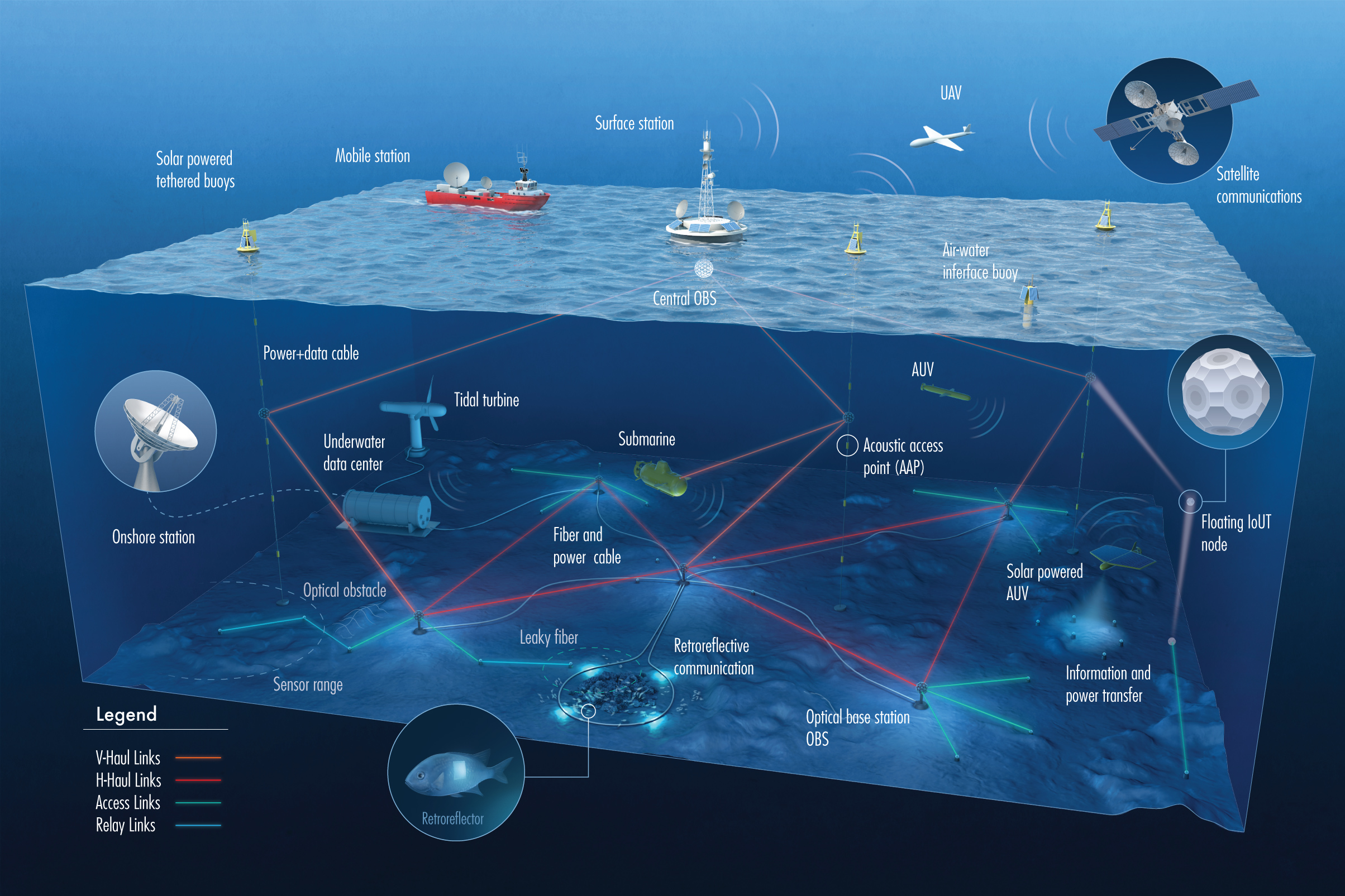}
\caption{The envisioned software-defined opto-acoustic network architecture.}
\label{fig:iout}
\end{center}
\end{figure*}

\section{An Opto-Acoustic Network Architecture}
UOANs can be designed either in an infrastructural, ad-hoc, or mixed fashion, as illustrated in Fig. \ref{fig:iout}. In the former, optical base stations (OBSs) and acoustic access points (AAPs) provide coverage to IoUT nodes in their vicinity and serve as a gateway to the rest of the network. In the latter, IoUT nodes are scattered across the network, and connectivity is established by the participation of IoUT nodes along a routing path, that is dynamically and distributively determined as per network conditions. In the mixed scenario, available infrastructure can be extended in both vertical and horizontal directions by using ad-hoc IoUT nodes. In what follows, we consider a mixed architecture in a layer-by-layer approach:

The \textit{perception layer} consists of sensors for data collection; actuators to interact with the physical world; and autonomous underwater vehicles (AUVs) to perform collaborative mobile tasks. Sensors can be designed to sense several physical phenomena (e.g., temperature, salinity, pollution, pH levels, etc.), store collected data in memory, and transmit observations to a target destination using UAC and/or UOC modems. IoUT nodes can also use passive transceivers such as acoustic tags, passive integrated transponders, optical retro-reflectors, which are especially suitable for cost-effective sensors with limited size and battery. On the other hand, AUVs are sophisticated platforms comprising of subsystems such as on-board computation and data processing; positioning-acquisitioning-and-tracking (PAT) mechanisms, navigation systems; and a variety of sensors/actuators. AUVs can also aggregate data from remote sensors using backscatter communication, as shown in Fig. \ref{fig:iout}. AUVs are able to operate at different depths based on hardware specifications and are generally powered by rechargeable batteries that can be charged by on-board solar panels.  

The \textit{network layer} is responsible for interconnection of all sub-systems as well as data collection, dissemination, and transportation. It mainly consists of access and transport networks. The access network comprises of links among perception layer and transport network components such as OBSs and AAPs. OBSs are in the shape of a multi-faceted sphere (i.e., geodesic polyhedral) and can provide omnidirectional connectivity by transceivers built at each face. Seabed OBSs can communicate with each other with fibers deployed in the seabed and/or horizontal collimated light-beams (i.e., H-Haul links). On the other hand, they can use vertical collimated light-beams (i.e., V-Haul links) to reach the central OBS via intermediate OBSs which are hanged on the tether of solar-powered buoys. While a tidal turbine can power seabed OBSs, the central and intermediate OBSs can be powered by solar panels of the surface station and tethered buoys, respectively. Solar-powered tethered buoys can also provide power and data to several AAPs down to the moor at the seabed. Air-water interface buoys may be used as an alternative solution. All these sea-surface systems connect IoUT nodes to the terrestrial IoT ecosystem via mobile stations, unmanned aerial vehicles (UAVs), and satellites. 

The \textit{middleware layer} provides a mediating language between components, interfaces, protocols, and IoUT applications. It is indeed an abstraction layer that masks the lower layer processes from the applications. Therefore, it is responsible for supporting efficient network management tools that can observe and manipulate the software and hardware features of the lower layers. The middleware should assure that communication pairs are identified, reachable, and ready to exchange data. Moreover, it can authenticate nodes for secure communication and establish an agreement about data integrity and error recovery. The middleware architecture can follow a service-oriented approach that decomposes the complex IoUT architecture into well-defined simpler subsystems operating on common interfaces and standard protocols such as routing and transport protocols, localization and tracking, energy efficiency,  security, and privacy.   

\begin{figure}[!t]
\begin{center}
\includegraphics[width=0.95 \columnwidth]{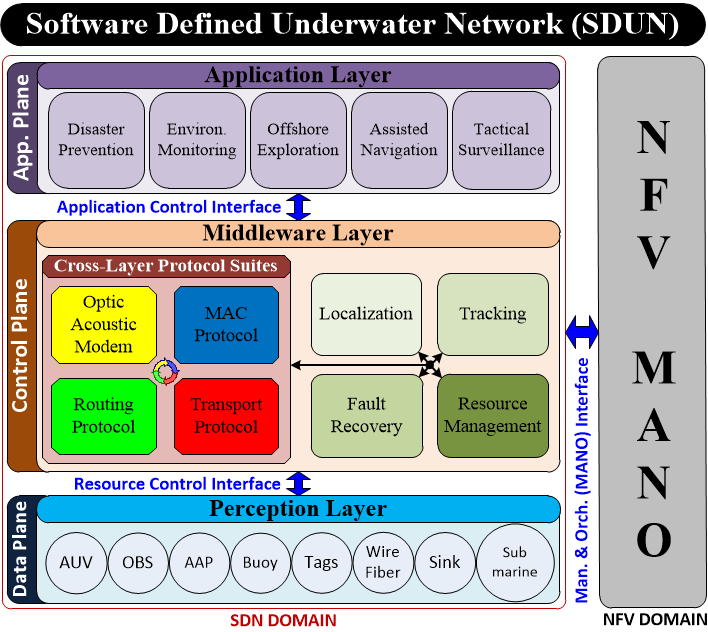}
\caption{A schematic diagram of integrated SDUN and NFV.}
\label{fig:SDN}
\end{center}
\vspace{-2.0em}
\end{figure}

The \textit{application layer} handles the final tasks of storing, organizing, processing and sharing the data aggregated from the lower layers. Execution of these tasks depends on application requirements which may differ from one to another. Unfortunately, existing underwater networks are generally designed for a specific application using this ossified layered architecture and built upon non-reconfigurable hardware-based proprietary equipment and services. As a remedy, software-defined IoUT nodes can enable a flexible UOAN architecture by reprogramming the hardware with an open and standardized interface (e.g., OpenFlow) that is compatible with multiple vendor-platforms. This paves the way for the novel SDM concept that characterizes the operation of network hardware by abstracting the control from data (infrastructure) plane [c.f. Fig. \ref{fig:SDN}]. Considering the heavy burden of upgrading the firmware by retrieving underwater equipment from deep waters, software-defined underwater network (SDUN) is indeed a radical paradigm shift for adapting to dynamic underwater environment and evolving to embrace emerging technologies \cite{AKYILDIZ20161}. Thanks to service virtualization support of cloud computation, network functions can be virtualized to provide flexibility, agility, and scalability as per the diverging needs of different applications. Hence, an NFV management and orchestration (MANO) system is required to govern virtualized infrastructure and SDUN. Indeed, MANO is an elegant solution to provide new services, maintaining existing services, and efficiently utilizing network resources. 

However, integrating SDUN and NFV technologies requires a centralized controller unit with significant computational power and data storage. Fortunately, these demands can be met with a tidal turbine-powered underwater data center (UDC), as shown in Fig. \ref{fig:iout}. Microsoft has recently launched project Natick (https://natick.research.microsoft.com) where a cylindrical tube-shaped UDC was sunk into the sea for the sake of significantly reduced cooling costs and offshore renewable energy sources. 

\section{SDUN+NFV: Enablers of Application Specific Cross-Layer Protocols}
In this section, we present recent advances and present challenges of UOANs. As a potential solution, we discuss how marriage of SDUN and NFV can enable cross-layer protocol suites tailored to the needs of different applications. 

\subsection{Perception Layer}
\paragraph*{Energy Shortage}
 IoUT nodes are expected to have a higher power budget to mitigate transmission loses and to compensate channel impediments by using sophisticated transceivers and signal processing. Considering the engineering challenges in charging and replacing batteries, energy self-sustainability driven lifetime maximization must be one of the leading design criteria. Indeed, microbial fuel cells and piezoelectric polymer strips can convert metabolic activities of bacteria and water flow energy into electrical power, respectively \cite{Celik2019Localization}. Alternatively, solar-powered AUVs may be used for simultaneous power and data transfer [Fig. \ref{fig:iout}]. By using a solar panel as a harvester and receiver, blue lasers have been shown to deliver around 7 mW power per minute and 0.5 Mbps data at the same time \cite{Trichili2019towards}. Even though lasers allow power transfer at longer distances, LEDs can provide a higher power rate since solar panels are designed to exploit entire visible light spectrum. 

\paragraph*{Hardware Limitation}
 In addition to the hardware complexity, the monetary cost of waterproof IoUT equipment may be high due to the need for extra protection from corrosion and fouling. Noting that commercially available low-cost acoustic modems have prices ranging from hundreds to thousand US dollars, mass production of optical modems can put more competitive price tags. As a result of the range-beamwidth tradeoff, light-emitting diodes (LEDs) can communicate with many surrounding nodes whereas a razor-sharp laser beam can reach distant users at desirable rates. For mobile nodes, lasers require effective PAT mechanisms which increase monetary cost, power consumption, and form factor of IoUT nodes. Hence, PAT should be considered only for AUVs rather than low-cost ad-hoc IoUT nodes. Similar to OBSs, ad-hoc IoUT nodes can also be designed as multifaceted transceivers to mitigate the need for PAT mechanisms by realizing omnidirectional communications. For instance, a truncated icosidodecahedron shaped node can host eight transmitters and receivers with a $\pi/4$ divergence angle and field-of-view (FoV), respectively.
 
\paragraph*{PHY \& MAC Issues}
Software-Defined Open-Flow enabled  IoUT nodes can overcome the above limitations by adapting physical (PHY) and medium access control (MAC) layer functions to the ever-changing underwater environment. As per the needs, they can prefer a suitable hardware front-end (e.g., acoustic or optic) and adjust the baseband processing parameters (e.g., modulation, coding, rate, power, divergence angle). For instance, the need for PAT mechanisms can be eliminated by adapting beamwidths to achieve robust and reliable links \cite{Celik2019Endtoend}. Frequency-division multiple access (FDMA) and time-division multiple access (TDMA) are not suitable for UACs because of the limited bandwidth and long-time guards, respectively. On the contrary, code-division multiple access (CDMA) is preferable thanks to its robustness against the multipath and frequency-selective fading effects. Indeed, the spectral efficiency of UAC can be further enhanced by using code-domain non-orthogonal multiple access (NOMA). Likewise, optical CDMA can improve UOC performance thanks to its spectral efficient and asynchronous nature \cite{Akhoundi}. UOCs can also benefit from optical schemes such as wavelength-division multiplexing (WDM). By combining of NOMA and WDM, NOMA can relieve multiple access interference by multiplexing several nodes into each wavelength provided by WDM. This combination is especially promising to serve a large number of nodes and realize different virtual networks on available network resources. 

 
\subsection{Network Layer}
The network layer tackles two types of traffic; control and data.  Thanks to omnidirectional and long-range communications, UAC links can ubiquitously deliver control signals to almost all destinations.  Although high-speed low-latency OWC links are more suitable to carry out the data traffic, they can also mitigate the distance-dependent capacity of UAC links. In this way, UANs and UONs complement one another to utilize the advantages of both. Nevertheless, multihop communication is still necessary to overcome high latency and short range of single-hop UAC and OWC links, respectively. Multihop communication can enhance the end-to-end (E2E) system performance by expanding the coverage area, extending the communication range, improving energy efficiency, and boosting network connectivity. 

\paragraph*{Routing}
The full benefits of multihop communication can only be gained with an effective routing algorithm that accounts for the underwater channel characteristics. Although there exist promising routing protocols proposed for UANs, they are not readily applicable for UONs due to the directivity of light-beams. In this regard, geographical routing protocols are especially suitable for UONs as location information is already needed for pointing between the transceivers. In \cite{Celik2019Endtoend}, we developed centralized routing protocols by tailoring shortest-path algorithms to different objectives (e.g., rate, error, power). Our investigations have shown that PAT has a significant impact on network performance, especially when location uncertainty and water turbidity increases. Unlike the conventional routing techniques that unicast packets to a single forwarder, opportunistic routing (OR) can reap the full benefits of broadcast nature of OWCs by targeting a set of candidate relays at each hop. The OR improves the packet delivery ratio as the likelihood of having at least one successful packet reception is much higher than that in unicast routing. Hence, OR is especially suitable for the underwater environment as the link connectivity can be disrupted easily due to the underwater channel impediments (e.g., pointing errors, misalignment, turbulence, etc.) and sea creatures passing through the transceivers’ line-of-sight. 

\begin{figure*}[t]
\centering
\begin{subfigure}[b]{0.32\textwidth}
\includegraphics[width=\columnwidth]{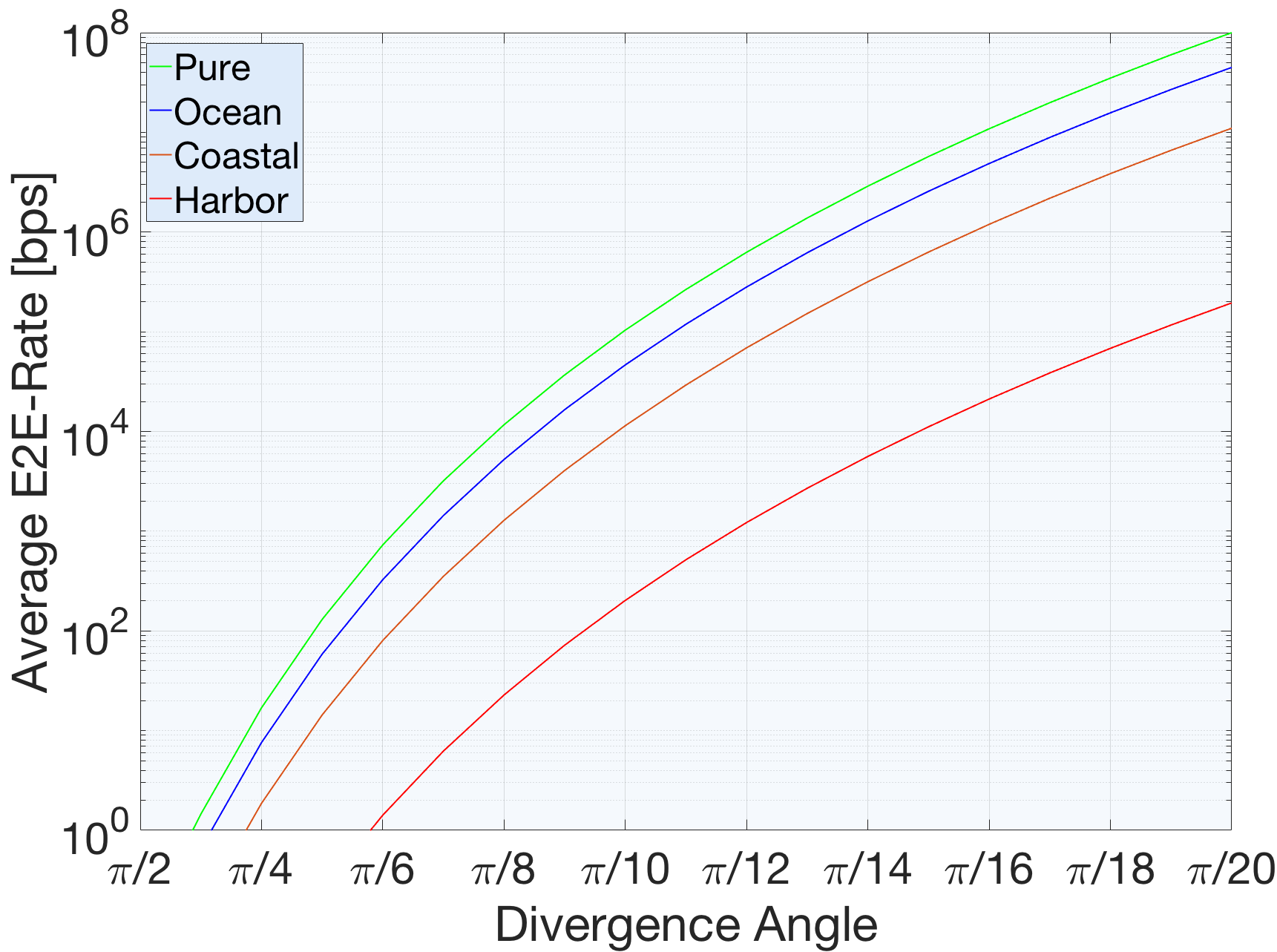}
\caption{}
\label{fig:rate}
\end{subfigure}
\begin{subfigure}[b]{0.32\textwidth}
\includegraphics[width=\columnwidth]{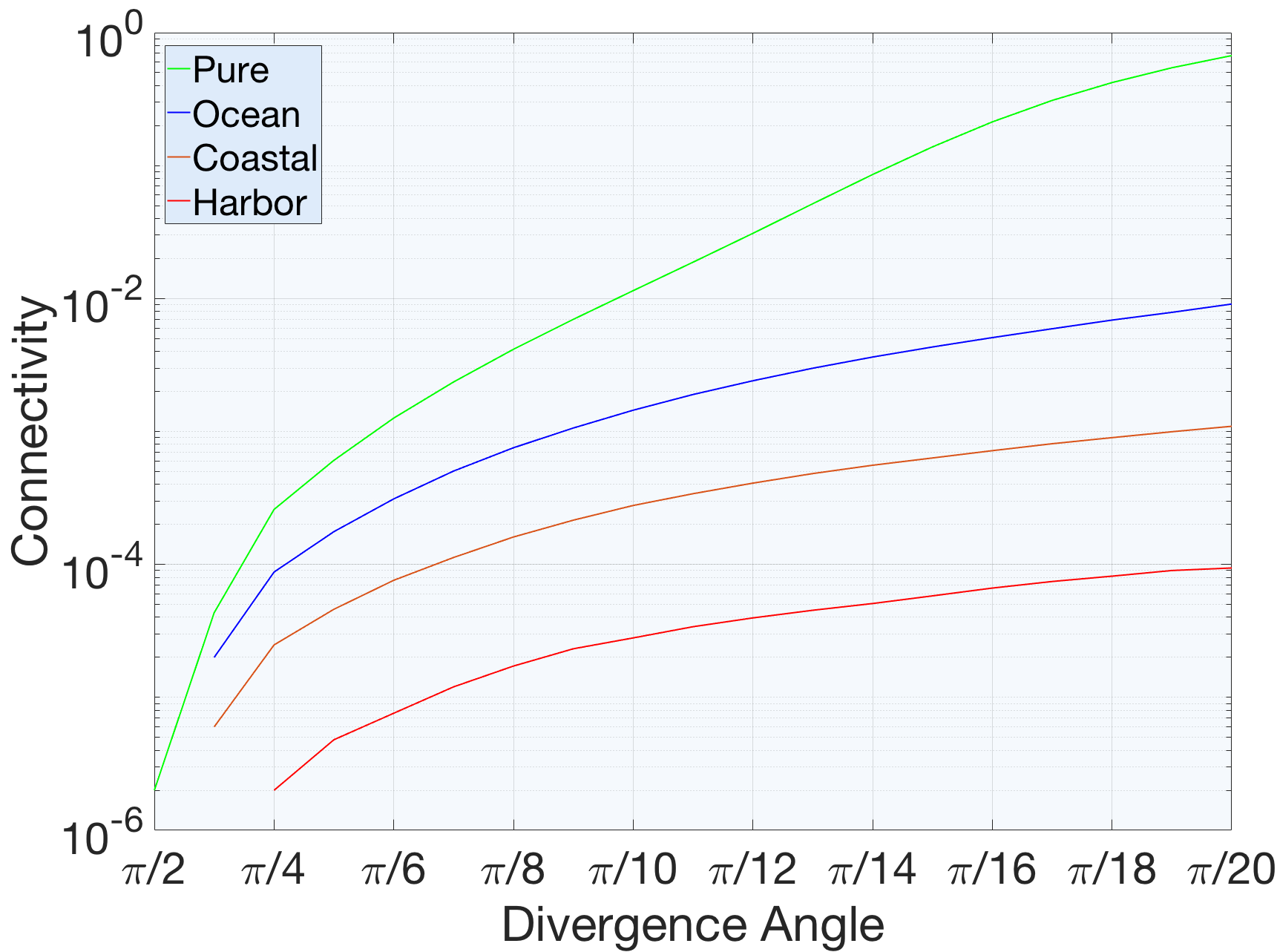}
  \caption{}
  \label{fig:connectivity}
\end{subfigure}
          \begin{subfigure}[b]{0.32\textwidth}
\includegraphics[width=\columnwidth]{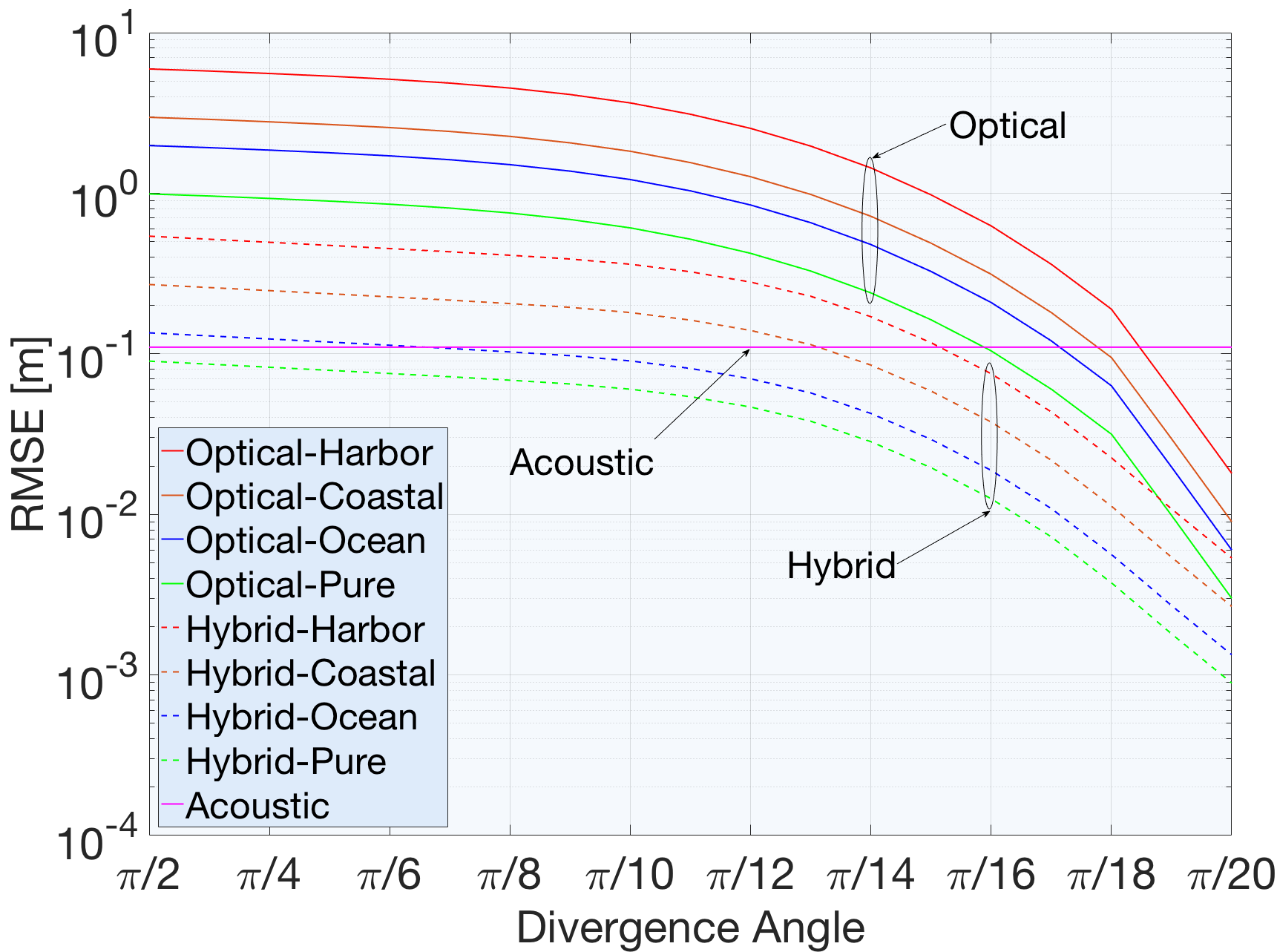}
  \caption{}
          \label{fig:loc}
\end{subfigure}
    \caption{The impact of divergence angle on network performance: a) E2E-Rate, b) Connectivity, and c) Localization error.}
                    \label{fig:faces}
\end{figure*}

\paragraph*{Connectivity}
As a result of the cost and deployment challenges, IoUT node density is expected to be sparse compared to that of IoT. Network density is a key determinant for the degree of connectivity that impacts many network performance metrics such as reliability, routing, capacity, and localization. Indeed, UANs can relieve node sparsity thanks to omnidirectional and long-range UACs. On the other hand, UON connectivity can drastically decrease with the node sparsity as PAT is not an available option for all nodes. To see the impacts of beamwidth and multifaceted IoUT nodes as a potential solution, let us consider an ad-hoc UON consisting of 50 nodes which are randomly distributed over a water volume of 500 $\rm{m}^3$. Following the Beer-Lambert channel model \cite{Celik2018Survey}, Fig. \ref{fig:rate} depicts the average E2E-Rate from IoUT nodes to a surface station located at the origin. By setting the bit error rate to forward error control threshold, the routing paths are determined to maximize the E2E capacity using the widest-path algorithm. Increasing the number of faces has a significant impact for two reasons: 1) A larger number of faces decreases divergence and FoV angles, which results in higher received power as well as concentrator gain; and  2) The incoming lightbeam can be acquired by multiple faces looking toward the transmitter, that naturally enhances the performance thanks to a higher reception diversity. In this way, multifaceted IoUT nodes can mitigate the link failures due to the pointing and alignment disruptions caused by random movements of sea surface or deep currents. Likewise, Fig. \ref{fig:connectivity} demonstrates the probability of connectivity that assures an E2E-Rate no less than 0.1 Mbps for all IoUT nodes. Fig. \ref{fig:faces} also shows the performance under different water types such that negative impacts of absorption and scattering effects that exacerbate with the turbidity of water.

\paragraph*{Reliability}
Ultra-reliable low-latency communications (URLLCs) require traffic to be forwarded with a limited packet loss at a predetermined rate. During transportation, packet losses may occur in shadow zones where temporary connectivity losses and high bit error rates are caused by severe channel conditions and/or network congestion. In such cases, a transport protocol is required to check corrupted data using error correction codes, order packets, verify the correct reception by ACK/NACK messages, and initiate retransmission lost packets. Transport protocol can be either connection-oriented, e.g., transmission control protocol (TCP), or connectionless, e.g., user datagram protocol (UDP). Even though the TCP considers the congestion as the only reason for packet loss, connection loss is quite common due to obstruction, pointing, and misalignment events. To distinguish between the two, connection-oriented underwater transport protocols should interact with the lower layers to leverage critical information. On the other side, as a simple connectionless protocol, the UDP cannot guarantee ordered and correct packet delivery. The minimum handshaking preference of the UDP exposes  IoUT nodes to the unreliability of the underwater networks. Hence, UDP is much more suitable for peer-to-peer transmission over quasi-stationary channel conditions such as transmission between nearby ground IoUT nodes or between AUVs and IoUT nodes [c.f. Fig. \ref{fig:iout}]. 

\paragraph*{Congestion}
Network congestion is caused by over-subscription of many flows to links with a capacity less than required. In such cases, routing and transport protocols must collaboratively adapt congestion window scale or flow rate to avoid buffer overrun along the path toward the destination. Since the sliding congestion window heavily depends on the precise estimation of round trip time (RTT), important TCP flow control mechanisms (e.g., timestamps to order packets and retransmission timeout) are prone to the high and variable delay of low-speed UACs. Even though this can be mitigated by low propagation and transmission delay of OWCs, transport and routing protocols should identify link and node failures caused by changes in channel states and network topology. Likewise, rate-adaptive schemes (e.g., proportional rate reduction) adjust  the transmission rate of the source to control the congestion by means of feedback control messages, which are also susceptible to long and highly variable RTTs. 

\subsection{Middleware Layer}
The layered approach was originally designed for rigid (non-programmable) infrastructure via isolation of network functions.  Such simplification of the network management yields a sub-optimal performance, especially in harsh and dynamic environments since these functions are indeed joint and heavily depend on each other. By virtue of integrating SDUN and NFV concepts, these closely intertwined associations among the layers can be captured and governed by MANO interface. The MANO can be thought of as a sophisticated cross-layer protocol suite that orchestrates network functions by collecting valuable information and making critical decisions. While MANO can allow IoUT nodes to locally manage PHY \& MAC functions to address rapid changes of the underwater environment, it should harmonize the cooperation among the higher layers, and then coordinate with nodes to instruct necessary local changes based on the global network view. In addition to behaving as a cross-layer protocol suite, the MANO should also provide the following services:  
 
 \paragraph*{Fault Recovery}
Since connectivity and reliability are the main delimiters of UOANs, predicting and handling shadow zones are of vital importance. Even a single node failure can result in significant performance degradation if this node plays the role of a gateway by connecting two partitions of nodes, which is not surprising in sparse networks. In case of failure, the nodes should be able to recover themselves by means of built-in self-organization and self-configuration algorithms. In the meanwhile, the MANO can handle ongoing traffic by steering it to an alternative routing path and help the faulty node to recover if necessary. Once the node recovers itself and a seamless connection is re-established, the traffic can be restored on the original path. At this point, hybridizing optic and acoustic networks comes into prominence as ubiquitous connectivity of UACs can help MANO to sustain ACK/NACK messaging between the nodes to recover from shadow zones. 

\paragraph*{Localization \& Tracking}
Localization is of utmost importance because of three reasons: 1) For many applications, collected data is beneficial only if it is tagged with an accurate position; 2) Geographical routing schemes can operate with location estimates; and 3) PAT mechanisms can work efficiently if the location of transceivers is known. In particular, network localization can provide the CPS with a general overview of the network, that can help to overcome congestion and connectivity loss. Unfortunately, the global positioning system (GPS) signals cannot propagate through the water due to the unique underwater channel characteristics. Noting that UAN localization literature is very well established, whereas recent UON localization schemes are designed based on either received signal strength (RSS) or time of arrival (ToA) based ranging techniques \cite{Celik2018Survey}.


Regardless of the underlying channel, localization accuracy relies upon quantity and deployment of surface anchors whose locations are obtained by GPS. Hence, GPS error propagates throughout the network and degrades the overall localization performance. As a result of strong winds and deep current, random movement of anchors and IoUT nodes cause additional uncertainty to the location estimates. At this juncture, we should also note that connectivity has a substantial impact on localization accuracy as having more measurements intuitively reduces the localization errors \cite{Celik2019Performance}. Therefore, hybridization of UANs and UONs can also improve the localization to a great extent. Let us revisit the previous network set up to observe the root mean square error (RMSE) of an RSS-based network localization with eight randomly located anchor nodes. Fig. \ref{fig:loc} shows that UAC delivers better performance than the UOCs due to the higher connectivity yielded from long-range omnidirectional communications. This behavior is reversed only after surpassing a divergence angle threshold as connectivity enhances with longer transmission ranges thanks to narrower beamwidth and FoV. Finally, the opto-acoustic case shows the positive impact towards a more precise localization, especially when UOC suffers from water turbidity. 

Furthermore, localization is of paramount importance for tracking algorithms which are needed for navigation, surveillance, AUV trajectory optimization, and animal monitoring. Even though tracking has conventionally been implemented by using acoustic tags, retro-reflective OWC can substantially improve efficiency. For instance, acoustically guided AUVs can also be identified by their retro-reflective tags whenever they pass nearby an optical IoUT node. Similarly, animal tracking can be realized by installing small-factor retro-reflective tags on fish. In particular, leaky fibers can be wired around a fish habitat for the purpose of simultaneous illumination and communication [c.f. Fig. \ref{fig:iout}]. 

\paragraph*{Resource Management}
The MANO has to manage two precious underwater resources; power and bandwidth. IoUT node lifetime can be significantly improved by energy-efficient PHY \& MAC functions such as power control, bandwidth allocation, and retransmission scheduling. More importantly, a sleeping strategy can jointly be designed with an opportunistic transmission scheme to buffer receiving traffic during the sleep and wake up for sending at desirable network and channel conditions to save energy. The local sleeping strategies should be regulated by MANO to maintain a seamless connection for incoming traffic requests. Accordingly, routing algorithms should discover routes by considering the available energy and buffer size of nodes along the path in coordination with lower and higher layer functions. 

\subsection{Application Layer}
With the advent of network softwarization, multiple virtual networks (a.k.a. network slices) can operate on a shared physical SDUN infrastructure. Network slices can be regarded as self-contained and mutually-isolated logical networks with their own tailored network functions and resources. Hence, diverging service demands of different applications can be met on the same infrastructure since each slice is independently orchestrated as per its own service requirements. These appealing features would incentivize multiple tenants to invest in building SDUNs in a cost efficient manner cooperatively and effectively monetize this investment by renting SDUN to proliferating IoUT applications.  

In addition to middleware layer functions mentioned above, the MANO must also manage the following application layer issues: 1) Quality-of-service-and-experience; 2) Service reliability; 3) Charging and billing; 4)  Network slice update/upgrade; and 5) VNF scaling, migration, and life-cycle management. In fact, developing these many VNFs for each application is not a practical approach. Instead, the network should be sliced into application classes for which VNFs should be developed by accounting for common service characteristics. For instance, acoustic part of the UOAN can be considered as a single slice and dedicated for control signaling of the MANO. Likewise, data traffic can be carried out over multiple slices which are allocated for different classes such as delay-tolerance, ultra-reliability, low-latency, high-bandwidth, etc. 
\section{Conclusions and Future Research Directions}
In this paper, we have shared our vision towards UOANs to overcome underwater environment hostility by reaping the full benefits of UANs and OANs. Instead of an ossified layered approach, we presented UOAN architecture in a cross-layer fashion and discussed entangled relations among different network functions.  By merging SDUN \& NFV domains, we also pointed out potential network virtualization-based solutions. In the remainder, we list some potential future research directions:

\paragraph*{Machine Learning}
Existing channel models are either simplistic by ignoring some crucial factors or analytically intractable due to complex formulations. Indeed, data-driven machine learning (ML) techniques are an indispensable tool to complement these traditional mathematical models by adjusting model parameters based on actual water characteristics. In terrestrial networks, ML has already received significant attention to select and optimize underlying physical layer functions such as modulation, coding, pointing, tracking. Power and bandwidth allocation can also be implemented based on ML tools by predicting dynamically changing channel conditions and traffic loads. Realizing all these features in a central unit yields a substantial communication overhead which may need for a notable portion of already limited network capacity. On the other hand, a fully distributed approach may place an intolerable burden on IoUT nodes with limited battery and computational power. Accordingly, a federated learning approach can strike a good balance between centralized and distributed solutions. In this case, IoUT nodes can act independently based on exchanging local and global network states with the central unit. Federated learning can also be extended to higher layers for improving routing, reliability, connectivity, fault tolerance. Therefore, ML can be the enabler of a self-organizing, self-optimizing, and self-healing SDUN architecture. 

\paragraph*{URLLC vs. Delay-Tolerant Communications}
URLLC applications pose great challenges on reliability and latency, which are generally conflicting design objectives. However, existing underwater research focuses on improving the capacity of the underwater networks with little attention to URLLC. For real-time applications, the network slice and related functions must be engineered well to satisfy underwater URLLC requirements. On the contrary,  some IoUT nodes may be located in remote areas for delay-tolerant applications. In this case, AUVs can play essential roles in collecting data and charging the IoUT nodes. Optimizing the AUV trajectory for maximum energy efficiency and data collection is an interesting research direction. Moreover, flying networks (e.g., UAVs, high altitude platforms) can provide global connectivity for surface elements (e.g., buoys, air-water interfaces, etc.) located outside of cellular coverage. In this case, jointly coordinating UAVs and AUVs would enable a more effective data aggregation.
 
\paragraph*{OBS/AAP Deployment and Controller Placement}
As the number of IoUT nodes and network size scales up, OBS/AAP deployment plays a crucial role in key network performance metrics. Deployment problem can be formulated to enhance delay, spectral and energy efficiency, and connectivity. Stationary OBSs/AAPs are also potential anchor nodes whose number and location may have a direct impact on localization accuracy. As they are powered with renewable energy, stationary OBSs/AAPs could also be used as distributed controllers to overcome the limitations on reaching to the central single-central UDC. Alternatively, AUVs may be used as mobile controllers to mitigate network topology and traffic pattern changes. Likewise, floating air-water interfaces might serve as surface anchors and controllers to improve overall network performance. Jointly optimizing number of controllers and their locations are challenging yet interesting problems to be investigated.  

\paragraph*{Integration with 5G and Beyond}
The lessons learned from merging SDN and NFV for 5G networks can be of great help to design the proposed IoUT network architecture. Since IoT is already provisioned as a slice of 5G networks, IoUT would be regarded as an extension to the IoT applications. In fact, many IoT applications (e.g., environmental monitoring, public health and safety, smart city, etc.) can work more effectively by data and inference provided by IoUT applications.

\bibliographystyle{IEEEtran}
\bibliography{IoUT}
\newpage
\begin{IEEEbiography}{Abdulkadir Celik}(S'14-M'16) received the Ph.D. in electrical and computer engineering from Iowa State University, Ames, IA, USA, in 2016. He is currently a postdoctoral fellow at CEMSE division of KAUST. 
\end{IEEEbiography}

\begin{IEEEbiography}{Nasir Saeed}(S'14-M'16-SM'19) received the Ph.D. degree in electronics and communication engineering from Hanyang University, Seoul, South Korea, in 2015. He is currently a postdoctoral fellow at CEMSE division of KAUST. 
\end{IEEEbiography}

\begin{IEEEbiography}{Basem Shihada}(SM'12) received the Ph.D. in computer science from University of Waterloo, Ontario, Canada in 2007.  He is currently an associate professor of CEMSE division at KAUST.
\end{IEEEbiography}

\begin{IEEEbiography}{Tareq Y. Al-Naffouri}(M'10-SM'18) received the Ph.D.  degree  in  electrical  engineering  from  Stanford  University,  Stanford,  CA, USA, in  2004. He is currently an associate professor of CEMSE division at KAUST.
\end{IEEEbiography}

\begin{IEEEbiography}{Mohamed-Slim Alouini} (S'94-M'98-SM'03-F'09) received the Ph.D. degree in Electrical Engineering from the California Institute of Technology, Pasadena, CA, USA, in 1998. He is currently a full professor of CEMSE division at KAUST.
\end{IEEEbiography}

\end{document}